\documentclass[10pt,english]{IEEEtran}
\usepackage[T1]{fontenc}
\usepackage[latin9]{inputenc}
\usepackage{geometry}
\usepackage{amstext}
\usepackage{graphicx}
\usepackage{babel}
\begin{document}

\title{Protection and Deception: Discovering Game Theory and Cyber Literacy
through a Novel Board Game Experience}

\author{Saboor Zahir, John Pak, Jatinder Singh, Jeffrey Pawlick, Quanyan
Zhu%
\thanks{Department of Electrical and Computer Engineering, Polytechnic School
of Engineering, New York University, Brooklyn, NY 11201. \{sz903,
jp3122, js6160, jpawlick, quanyan.zhu\} @nyu.edu

This work was supported in part by the New York University Prototyping
Fund, a collaborative program offered by the \emph{Greenhouse at NYU}
and the \emph{NYU Entrepreneurial Insitute}.

It was also supported in part by an NSF IGERT grant through the \emph{Center
for Interdisciplinary Studies in Security and Privacy} (\emph{CRISSP})
at NYU.%
}}
\maketitle
\begin{abstract}
Cyber literacy merits serious research attention because it addresses
a confluence of specialization and generalization; cybersecurity is
often conceived of as approachable only by a technological \emph{intelligentsia},
yet its interdependent nature demands education for a broad population.
Therefore, educational tools should lead participants to discover
technical knowledge in an accessible and attractive framework. In
this paper, we present \emph{Protection and Deception} (\emph{P\&G}),
a novel two-player board game. \emph{P\&G} has three main contributions.
First, it builds cyber literacy by giving participants ``hands-on\textquotedblright{}
experience with game pieces that have the capabilities of cyber-attacks
such as worms, masquerading attacks/spoofs, replay attacks, and Trojans.
Second, \emph{P\&G} teaches the important game-theoretic concepts
of asymmetric information and resource allocation implicitly and non-obtrusively
through its game play. Finally, it strives for the important objective
of security education for underrepresented minorities and people without
explicit technical experience. We tested \emph{P\&G} at a community
center in Manhattan with middle- and high school students, and observed
enjoyment and increased cyber literacy along with suggestions for
improvement of the game. Together with these results, our paper also
presents images of the attractive board design and 3D printed game
pieces, together with a Monte-Carlo analysis that we used to ensure
a balanced gaming experience. \end{abstract}


\section{Introduction\label{sec:Introduction}}

Cybersecurity has been directly in the limelight of contemporary media.
The Sony Pictures Entertainment hack over the controversial film \emph{The
Interview}, the infamous debut of the Snowden Revelations and ensuing
debate, and important security breaches at The Home Depot and Target
Corporation have made national news at all levels of society. The
U.S. Federal Government's commissions of reports on big data and privacy
\cite{key-2} and bulk collection of signals intelligence \cite{key-4}
- together with the surging interest in cybersecurity from academic
and commercial perspectives - suggests an intense effort to combat
cybercrime from the top-down. But cybersecurity is an interdependent
phenomenon. This interdependency demands cyber literacy that branches
out from technology companies and computer science schools to consumers
of the technology that they develop. It also requires a grassroots
effort at igniting interest in cyber-careers as an investment in tomorrow's
human capital. 

\emph{Serious games} offer a promising means to overcome the intimidating
nature of learning about cybersecurity. Because it is difficult to
perceive how security threats affect individuals, and because cyber
experience and vocabulary are not well-integrated among those in non-technical
fields, cybersecurity can seem to pose a high barrier to entry \cite{key-3}.
Serious games employ the entertainment value of games towards accomplishing
distinct educational objectives. They sit upon an intersection between
engineering, science, and education. Our work is a serious game with
the objectives of answering such basic questions as ``What is a masquerading
attack?'' and ``How is a local area network different from the internet?''

Several recent educational efforts have promise for technical professionals
or aspiring STEM students. Proliferating \emph{Capture the Flag} (\emph{CTF})
competitions have placed security education in a non-technical environment.
An application of \emph{gamification}, they leverage the enjoyable
properties of games in a real-life security challenge. But they may
not be appropriate for novice participants. They do not (at least
yet) especially represent an outreach of security education beyond
the STEM fields and into populations underrepresented in technical
fields. Games are needed that feature a gentle introduction to cyber-security;
one that helps build cyber literacy without intimidation and teaches
other concepts relevant to cybersecurity only implicitly.

In this work, we present \emph{Protection and Deception} (\emph{P\&G}),
a two-player board game that combines a turn-based chess-like structure
with elements such as infrastructure configuration that are characteristic
of real-time strategy games. The basic gameplay is simple and follows
a storyline related to cybersecurity. In \emph{P\&G}, both players
configure local area networks (LANs) and allocate attack and defense
packages. They hide ``critical information'' on one of their computers.
Then, gameplay evolves in a sequence of turns in which players deploy
attacks and navigate them through the network. Throughout the game,
players learn about attack capabilities. They also face trade-offs
between brute strength and maintaining information-assymetry - as
when deciding whether to surveil an opponent's LAN with a weak attack.
Players achieve victory when they destroy the opponent's computer
containing the critical information.

\emph{P\&G} offers a gentle introduction to cyber literacy. Explicit
cyber-jargon is limited to various types of cyber attacks: \emph{e.g.},
viruses, Trojans, masquerading attacks and worms. The rest of the
gameplay has parallels in traditional board games - although there
are some parallels to collectable card games (\emph{e.g. Magic: The
Gathering} and \emph{Yu-Gi-Oh!}). In this way, \emph{P\&G} attempts
to lower the learning curve for serious security games so that they
can reach populations outside of corporations or the university.

Indeed, we tested this game at a community center on the lower-east
side of Manhattan. We found both encouraging results - in terms of
interest in the game and acquired knowledge - and elements of the
game that need to be improved and further simplified in order to attract
young players. We were also inspired towards future work in digitalizing
the game or providing game instructions in the form of a \emph{YouTube}
video.

The rest of the paper proceeds as follows. Section \ref{sec:Gameplay}
describes the gameplay of \emph{P\&G} in detail. We were especially
intrigued by one aspect of the gameplay design: attempting to balance
the capabilities of cyber-attacks and defense packages. Towards this
end, we created a Monte-Carlo simulation which we describe in Section
\ref{sec:Simulation-for-Design}. Section \ref{sec:Playtesting} describes
our playtesting proceedure and observations. Finally, we conclude
the paper in Section \ref{sec:Conclusions}.

\section{Gameplay\label{sec:Gameplay}}

\emph{Protection and Deception }(\emph{P\&G}) is a two-player board
game. The goal is of the game is to locate and destroy the opponent's
computer that holds his \emph{critical information}. This task is
achieved through a combination of effective local area network (LAN)
design, intelligent deployment of attacks and defenses, and quickly
routing attacks and defenses to their intended target. This requires
balancing strength with maintaining the ability to deceive the other
player. The first stage of \emph{P\&G} consists of LAN configuration.

\subsection{LAN Configuration}

\begin{figure}
\begin{centering}
\includegraphics[width=0.8\columnwidth]{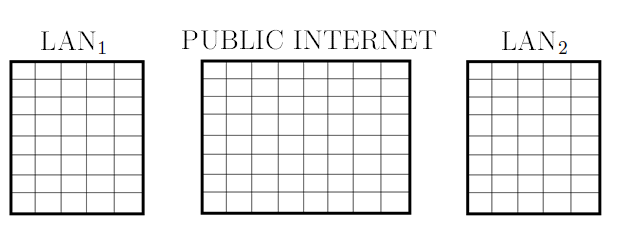}
\par\end{centering}

\protect\caption{\label{fig:Basic-board-layout.}Basic board layout. Each player configures
her own local area network, and connects via routers to the public
internet. Network configuration is an optimization experience in which
players need to make trade-offs between capabilities to defend critical
information, utilize deception (hide critical information in unexpected
computers), and rapidly deploy attacks.}

\end{figure}

Each player controls the following pieces: 
\begin{enumerate}
\item 4 routers
\item 8 computers
\item 8 mesh points
\item 16 links
\item Deck of attack and defense cards
\end{enumerate}
There are three components to the platform of the board game as shown
in Fig. \ref{fig:Basic-board-layout.}.

Each player will have a Local Area Network (LAN), which is essentially
her base. The players configure these LANs. The board that is in the
middle of the two LANs is the public Internet, which has static configuration.

\begin{figure}
\begin{centering}
\includegraphics[width=0.8\columnwidth]{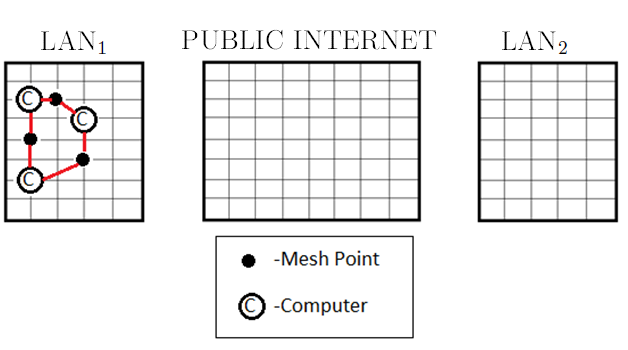}
\par\end{centering}

\protect\caption{\label{fig:Basic-board-layout2}Basic board layout with three computers
connected via mesh points and links}

\end{figure}

The game begins with each player setting up her own LAN. A network
topology consists of routers, computers, mesh points, and links. A
mesh point is essentially a way to link two computers directly. Fig.
\ref{fig:Basic-board-layout2} represents a sample network topology
for Player A of three computers connected with the use of three mesh
points.

\begin{figure}
\begin{centering}
\includegraphics[width=0.8\columnwidth]{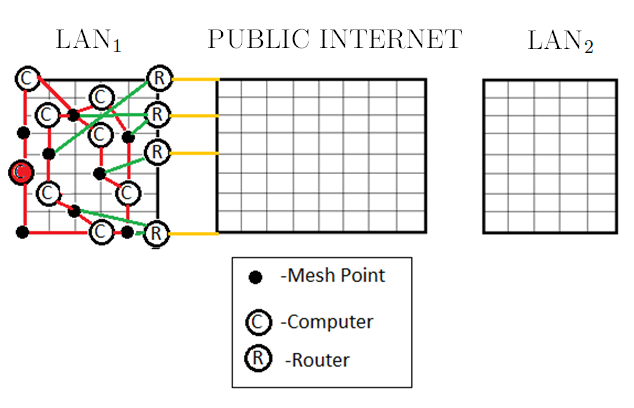}
\par\end{centering}

\protect\caption{\label{fig:Sample-configuration-of}Sample configuration of $\text{LAN}_{1}$.
The red computer holds critical information. The routing links are
represented with green lines, while the computers are linked to each
other or to mesh points with red lines.}

\end{figure}

Each player creates a network topology that consists of 8 computers,
at least 4 mesh points and at most 8 mesh points, and 4 routers, which
are accompanied by 4 routing links and must be connected to at least
four mesh points. A router is used to connect computers to the public
Internet. The 16 links are used in order to connect a computer to
a computer, a computer to a mesh point, or a router to a mesh point.
Each computer must use 2 to 3 links to connect to another computer
or mesh point. Fig. \ref{fig:Sample-configuration-of}  is a sample
network topology. The routing links are represented with the green
lines. In Fig. \ref{fig:Sample-configuration-of}, the computers are
linked to other computers or mesh points with the use of red links.
After each player sets up his or her network topology, each player
must designate one computer out of the eight computers as the computer
that holds \textquotedblleft critical information\textquotedblright .
In Fig. \ref{fig:Sample-configuration-of}, the computer that holds
the critical information is colored in red.

Each router is connected to the public Internet with the use one link
(in yellow in Fig. \ref{fig:Sample-configuration-of}). Once each
player has set up her network topology and decided on the computer
that holds \textquotedblleft critical information,\textquotedblright{}
each player allocates the deck of attack and defense cards. Each computer,
besides the computer that holds \textquotedblleft critical information,\textquotedblright{}
is assigned one attack card and one defense card. The next two subsections
describe the attacks and the defense packages.

\subsection{Attacks}

\begin{figure}
\begin{centering}
\includegraphics[width=0.8\columnwidth]{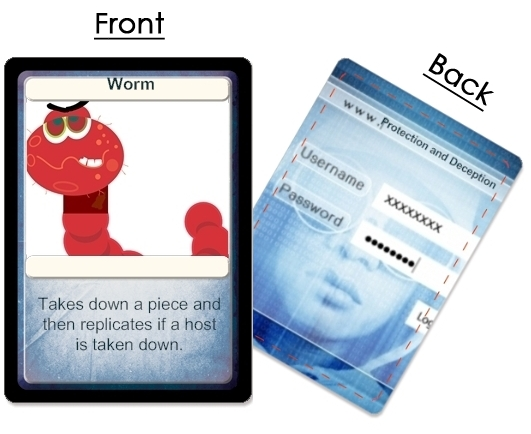}
\par\end{centering}

\protect\caption{\label{fig:Sample-attack-card:}Sample attack card: \emph{Worm} Attack}

\end{figure}

Each attack card features a different type of attack. These attacks
can be spawned from the computers equipped with the attack card. Fig.
\ref{fig:Sample-attack-card:} depicts an image of one of the attack
cards.
\begin{enumerate}
\item \emph{Worm} - takes down a piece and then replicates if a host is
taken down
\item \emph{Masquerading Attack/Spoof} - propagates throughout a network
without attacking a particular piece
\item \emph{Denial of Service} (\emph{DOS}) \emph{Attack} - stops traffic
within a mesh point. 
\item \emph{Virus} - attacks a computer and then the piece is reset
\item \emph{Replay} - captures a packet and does not let it propagate throughout
the network
\item \emph{Trojan} - reveals the defenses that a particular computer has
installed. A Trojan is also coupled with a weak level virus
\item \emph{Modification Message} - changes the type of message/attack a
computer sends. If this attack comes across the opponent\textquoteright s
attack at a node in the Internet, it can randomly select a different
type of attack
\end{enumerate}

\subsection{Defense Packages}

\begin{figure}
\begin{centering}
\includegraphics[width=0.8\columnwidth]{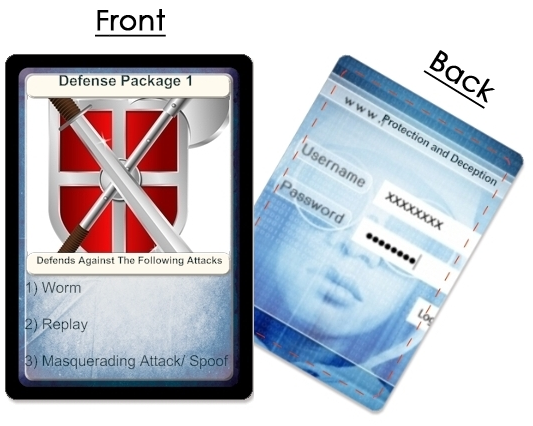}
\par\end{centering}

\protect\caption{\label{fig:Sample-defense-package}Sample defense package card: \emph{Defense
Package 1}}

\end{figure}

Players also equip computers with a defense package. The defense packages
differ in terms of which attacks they block. We have counterbalanced
these packages in order to prevent any one attack from becoming exceptionally
powerful. (See Section \ref{sec:Simulation-for-Design}.) Fig. \ref{fig:Sample-defense-package}
depicts one of the defense package cards.
\begin{enumerate}
\item \emph{Defense Package 1} - Blocks worm, replay, and masquerading attack/spoof
\item \emph{Defense Package 2} - Blocks worm, denial of service (DOS), and
modification message attacks
\item \emph{Defense Package 3} - Blocks worm, virus, and Trojan attacks
\item \emph{Defense Package 4} - Blocks worm, modification message, and
masquerading attack/spoof
\item \emph{Defense Package 5} - Blocks worm, Trojan, and DOS attacks
\item \emph{Defense Package 6} - Blocks virus, replay, and masquerading
attack/spoof
\item \emph{Defense Package 7} - Blocks Trojan, replay, and DOS attacks
\item \emph{Defense Package 8} - Blocks Trojan, replay, and modification
message attacks
\end{enumerate}

On every turn, each player is allowed to make one move. A move is
defined as either spawning an attack or moving an attack one unit.
An attack piece is represented as a ring. When a player spawns an
attack piece, she simply places the ring on top of the appropriate
computer. An attack that is in a LAN follows the configured links.
An attack that is in the public Internet moves along the sides of
the squares. The attack is allowed to move either horizontally (left
or right) or vertically (up or down). Each player does not know what
the other player\textquoteright s moving attacking is. An attack is
revealed under one of two conditions: 
\begin{enumerate}
\item A player attacks the opponent's attack
\item A player attacks the opponent's computer
\end{enumerate}
A defense package is revealed if an attack is conducted on a computer.
Below is a sample gameplay: 
\begin{enumerate}
\item Player A has a worm attack. Player A attacks Player B\textquoteright s
computer. 
\item Player B reveals the Defense Package that is assigned to the particular
computer that is attacked: Defense Package 1. 
\item Defense Package 1 is able to defend against a Worm, Replay, and Masquerading
Attack/Spoof. Therefore, Player A\textquoteright s worm attack is
destroyed. 
\end{enumerate}
If an attack attacks a computer and the computer is successfully able
to defend against the attack, the attack is destroyed. However, although
the attack is destroyed, it can still be spawned from the starting
point, which is the computer that the attack originated from, on another
turn. If an attack attacks a computer and the computer is unable to
defend against the attack, the computer is destroyed. The game ends
once one player discovers and destroys the opponent\textquoteright s
computer that holds the \textquotedblleft critical information\textquotedblright .

\section{Simulation and Strategy\label{sec:Simulation-for-Design}}

Every game requires fairness for a balance of good gameplay. No single
attack should dominate to the point where the game ends quickly. In
this section, we first describe the results of a simulation that we
used to balance the capabilities of the attacks and the defense packages,
and then we describe a strategy that might be employed based on insight
from this simulation design process.

\subsection{Simulation for Design}

\emph{Flow} is a notion developed by psychologists to describe a mental
state in which one is completely involved in an activity for its own
sake {[}1{]}. It is characterized as an activity where time flies.
Fairness in a game is essential to induce flow \cite{key-6}. We wanted
to allocate defense capabilities such that no single attack was able
to dominate. In order to do this, we simulated virus and worm attacks
against a fixed network topology for different allocations of defense
packages%
\footnote{We simulated virus and worm attacks because they have least and most
powerful special attack properties, respectively. The virus has no
special attack power, while the worm has the power to continue to
propagate if it is not destroyed. We allocated defenses against the
other attacks by assuming that their special attack properties lie
somewhere between those of the virus and worm. Thus, we configured
between two defense packages (the number which were endowed with the
ability to block viruses) and five defense packages (the number configured
with the ability to block worms) with the ability to block the other
attacks.%
}. This gave us a mapping from (number of defenses with the ability
to block viruses) to (number of computers that a virus would likely
destroy), and it gave us a similar mapping for worms. We then used
the inverse of this mapping to allocate the capabilities of defense
packages such that viruses and worms would be likely to destroy the
same number of computers.

\begin{figure}
\begin{centering}
\includegraphics[width=0.8\columnwidth]{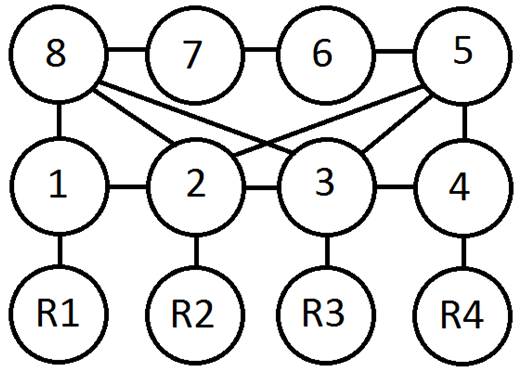}
\par\end{centering}

\protect\caption{\label{fig:LAN-topology-used}LAN topology used for Monte-Carlo simulation.
In this simulation, we ran different attacks against different possible
configurations of the defense packages. We ultimately designed the
defense packages based on the configurations that created the most
equal performance for the different attacks.}

\end{figure}

For the Monte Carlo simulation, we used the following topology in
Fig. \ref{fig:LAN-topology-used}, one that is within limits and is
symmetrical in nature. 

\begin{figure}
\begin{centering}
\includegraphics[width=0.8\columnwidth]{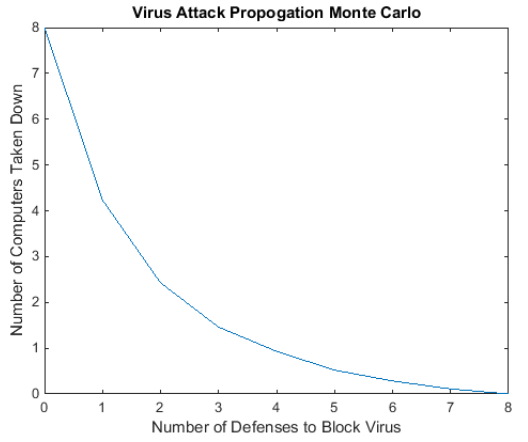}
\par\end{centering}

\protect\caption{\label{fig:Number-of-computers-destroyed}Number of computers destroyed
by virus attack versus number of defenses equipped with the ability
to block the virus. For instance, if four defenses were to be configured
with the capability to block the virus, then the average virus attack
would destroy approximately one computer.}

\end{figure}

\begin{figure}
\begin{centering}
\includegraphics[width=0.8\columnwidth]{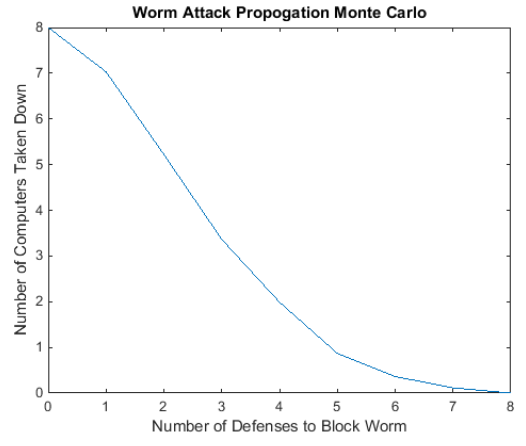}
\par\end{centering}

\protect\caption{\label{fig:Number-of-computers-worm}Number of computers destroyed
by worm attack versus number of defenses equipped with the ability
to block the worm}

\end{figure}

For the random simulations, a random routing point was chosen from
a uniform distribution. The virus was simulated such that it would
not revisit nodes if it had the potential to explore unvisited nodes.
The worm had the capability to visit all nodes. For each number of
defenses ranging from 0 to 8, 1000 simulations were done to average
the number of nodes destroyed. Figs. \ref{fig:Number-of-computers-destroyed}
and \ref{fig:Number-of-computers-worm} depict the results of these
simulations.

Figs. \ref{fig:Number-of-computers-destroyed} and \ref{fig:Number-of-computers-worm}
show that to give the virus and worm similar strengths, the number
of defenses that protect against viruses should be less than that
of worms. Based on the figures, four defense packages should be equipped
with the ability to block worms and two with the ability to block
viruses. This makes each able to destroy approximately 2.5 computers
on average%
\footnote{The first iteration of defense packages used preliminary simulation
results. Thus, in the allocations discussed in Section \ref{sec:Gameplay},
there are five rather than four defense packages equipped with the
ability to block worms. %
}.

Based on the results of this simulation, the next subsection describes
a sample strategic consideration that players might use to build a
LAN and allocate defense packages.

\subsection{Strategy}

Clearly there are some implicit guidelines for making a topology.
For instance, it seems unwise to leave a direct path without worm
defense to the critical computer. Such topologies arise in automatic
wins if the correct attack is carried out. One particular defensive
strategy that could be used is to create two \emph{communities}. 

\begin{figure}
\begin{centering}
\includegraphics[width=0.8\columnwidth]{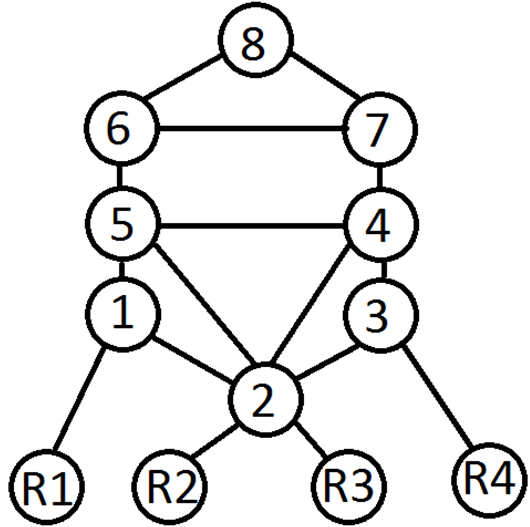}
\par\end{centering}

\protect\caption{\label{fig:A-topology-robust}A topology to robustly protect against
worm attacks}

\end{figure}
The topology in Fig. \ref{fig:A-topology-robust} is an example of
a dual-community topology. A community could be defined as a concentration
of nodes with a high degree of inter-connectivity. This dual-community
topology in Fig. \ref{fig:A-topology-robust} also has the property
that it forces attacks through certain computers on the way to computer
number 8, in which the critical information is maintained. As a result,
there are no short routes to get to node 8. 

\begin{figure}
\begin{centering}
\includegraphics[width=0.8\columnwidth]{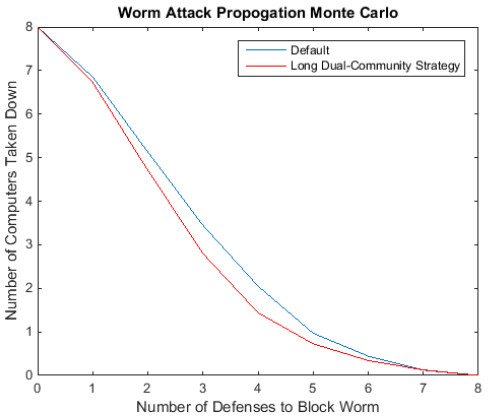}
\par\end{centering}

\protect\caption{\label{fig:Computer-casualties-with}Computer casualties with default
versus robust defense configurations}

\end{figure}

We conducted a simulation to analyze the effectiveness of this topology.
The results of this simulation are shown in Fig. \ref{fig:Computer-casualties-with},
which shows that fewer computers were eliminated on average for the
same defensive configurations for the long dual-community strategy
than for the default strategy.

From an offensive standpoint, an attack strategy might be to send
out 4 attacks simultaneously. The attack that has the least probability
of being defended against will attack a node. Once this node is attacked
the defenses of that node are now known. If it successfully defends
against one attack, the player has at least one other attack to take
out this node. In fact, this method of attack is very effective when
using the first attacker to be a virus because there are only two
defenses against viruses.

\subsection{Implicit Game-theoretic trade-offs}

Besides explicitly teaching players basic cyber literacy, \emph{P\&G}
also aims to give them implicit experience in game-theoretic optimization.
This optimization is apparent in network configuration for defense
and selecting optimal attack strategies.

The defensive network strategy embodied by the dual-community strategy
depicted in Fig. \ref{fig:A-topology-robust}, for instance, involves
a trade-off. The advantage of the configuration is that it strongly
protects computer number 8, which can be used to store the critical
information. Unfortunately, this also reveals the likely location
of the critical information to the opposing player! A more ``flat''
and network topology would have the advantage of more effectively
disguising the location of the critical information. Such deception
is heavily studied in the area of security in general \cite{key-13,key-14,key-15},
and is especially important in cybersecurity \cite{key-16,key-17}.
In terms of game theory, choosing a flat topology amounts to preferring
information asymmetry to brute force.

Information asymmetry is also important in selecting attack strategies.
Initially, an attacking player has no knowledge of her opponent's
allocation of defensive packages. She has the option to use initial
attacks primarily as ``scouts'' in order to ascertain the allocation
of defense packages. Of course, this may involve sacrificing the turns
that it takes to regenerate attacks. We are excited to see how players
develop strategies that leverage these concepts - possibly without
explicit knowledge of the scholarship behind them.

\section{Playtesting\label{sec:Playtesting}}

The initial target audience of \emph{Protection and Deception} (\emph{P\&G})\emph{
}was any person over the age of six. The game was tested out among
various ages ranging from ages six to 21 years old. The testers were
from two groups. The first was a combination of children who frequented
a community center located in the Lower East Side of Manhattan, New
York. The second consisted of mostly college students. Our initial
tests at the community center were conducted with four children.

We initially considered implementing structured pre- and post-play
surveys that would have enabled statistical analysis. Encouraged,
however, by advice from the educational community, we eventually opted
for less structured observation that would not discourage students.
Essentially, we collected evidence by open-ended observation. 

Questions to the children before the game lasted no more than five
minutes per player. We asked the players their age, what they know
about cyber security, what academic subject they preferred, and what
interests they pursued outside of school. The questions about favorite
subject and interest were a means to figure out their backgrounds.
We had children who were interested in math, science, basketball,
painting, and other activities. These children at the community center
did not have any explicit knowledge about cybersecurity. We asked
whether they had heard of ``hackers,'' but they had not. We also
tested the game with two high school students, one of which expressed
interest in business and another in engineering. Finally, our second
pool of testers were college students looking to pursue careers in
the fields of engineering, medical, and art. 

In the post-survey, all players were asked what they learned about
cyber security, what they liked and disliked about the game. One 7-year
old girl from the community center said, \textquotedblleft I like
the cards the most.\textquotedblright{} A 10-year old boy said, \textquotedblleft I
forgot which card I put down for the different computers\textquotedblright{}
- which indicated to us an aspect of the board design that we can
improve so that it is obvious which attack and defense cards have
been allocated to each computer. A 20-year old college student studying
medicine had a brief understanding about cybersecurity before the
game, but after playing \textquotedblleft learned how different attacks
such as the worm worked and learned about cyber attacks that I didn't\textquoteright t
know existed like masquerading.\textquotedblright{} A 17 year old
in high school who expressed an interest in business said he would
play the game if more of his friends knew about the game and how to
play. He was asked a follow-up question if there was anything in the
game he wanted to learn more about. He said he plays video games on
his\emph{ PlayStation 4} console a lot and realized he \textquotedblleft had
an experience of denial of service when a group of hackers took down
the \emph{PlayStation} online network and I could not log on or use
the network for a few days.\textquotedblright{} We were encouraged
by this rather comical realization that cybersecurity concepts are
especially embedded in non-academic activities.

For all age groups, the instructions seemed rather complex; many times
during the game, players would ask the testers whether moves were
legal or ask about the results of particular actions. Importantly,
we learned that it was helpful to follow the instructions with a quick
demonstration of the game play. This adjustment in our introduction
of the game decreased the difficulty of learning, although did not
remove the learning curve completely. Based on this expressed difficulty,
we are considering including video instruction or other means to make
the game easier to learn. We describe these briefly in Section \ref{sec:Conclusions}.

Finally, we noted that the game seemed enjoyable to players once the
rules became clear. Players were excited when their attack successfully
destroyed a computer or when their computers successfully repelled
the opponent's attacks. Among the older players, we noted a competitiveness
that emerged from the freedom allowed to choose different strategies.
From observing the various age groups, it appeared that the testers
that were around or over the age of 13 enjoyed the game the most.
We will seek a much larger subject pool for further testing in order
to refine the target age for \emph{P\&G}.

\section{Related Work}

In the introduction, we described various classes of games from which
\emph{Protection and Deception} (\emph{P\&G}) derives its framework.
Namely, \emph{P\&G }is a serious game - a game which teaches concepts
which have actual value outside of serving the entertainment purpose
of the game. \emph{P\&G} aims to build cyber literacy, as well as
to implicitly teach about trade-offs between strength and information
revelation. Furthermore, \emph{P\&G} represents an effort in the vast
category of security education, a critical area of study in the light
of intense regional and international conflicts in cybersecurity.
Finally, \emph{P\&G} builds upon a tradition of games-based learning.

We can see similarities to \emph{P\&G} in at several recent games.
From last year's \emph{3GSE, }Microsoft's \emph{Elevation of Privilege}
\cite{key-7} is a card game based on concepts from information security
with a fascinating purpose: it is played between developers in order
to discover security flaws of a system. \emph{Elevation of Privilege}
is an example of gamification, since it employs motivations from game-playing
for a serious task. Developers in this game draw cards which prompt
them to name vulnerabilities, and thereby accomplish a technical objective.
\emph{Elevation of Privilege }is obviously not geared towards a novice
population.

\emph{Control-Alt-Hack} \cite{key-8} is a card game from \emph{3GSE'14}
which is geared towards a novice population. This game seeks to give
participants an social experience related to hacking, rather than
teaching specific concepts. The network security game called \emph{{[}d0x3d!{]}}
\cite{key-9} is also a similar effort to ours. It is a board game
with changeable configuration achieved by tiles which are arranged
at the beginning of gameplay. Players in \emph{{[}d0x3d!{]}} deploy
special abilities on their way to collecting digital resources (``\emph{{[}loot{]}}'').
Both games are attractively designed, and represent efforts to intelligently
deploy and commercialize or test security games. They both use existing
games re-skinned in cybersecurity concepts and terminology, whereas
our game is an entirely new design.

\emph{Control-Alt-Hack} and \emph{{[}d0x3d!{]}} both seem to feature
a higher degree of security vocabulary than \emph{P\&G}. Indeed, \emph{P\&G}
represents an effort to reach out to non-technical, underrepresented,
and young players. We are concerned not only about players who may
not have the technological background to understand security concepts,
but also players who may not have the attention span to learn a complicated
game. In our own playtesting, we observed that even with the simple
mechanics of our game, there was some learning curve. Thus, we aim
to keep the security lexicon in \emph{P\&G} to a minimum. This will
help us achieve the goal of engaging a diverse population in security
awareness.

\section{Conclusions and Future Work\label{sec:Conclusions}}

Our initial work on \emph{Protection and Deception} (\emph{P\&G})
opens up vast possibilities for future development. In terms of basic
elements of gameplay, we have considered several options. First of
all, the connection between security challenges and economic questions
has been extensively noted in the literature \cite{key-10,key-11,key-12}.
Because of this, we are considering incorporating money or budgeting
resources into the gameplay. We are also considering allowing players
to elect to build up their LAN capabilities instead of deploying attacks.
This trade-off pits myopic against farsighted strategies, and allows
implicitly teaching the present value of future rewards. Finally,
we have noted that a visual demonstration of play seemed to lower
the learning curve for our participants. Because of this, we are considering
deploying video instructions online that can be used to learn the
game. On the more extreme end, the entire game could be digitized,
or a hybrid board game and digital game combination could be considered.

In its present version, \emph{P\&G} is a board game designed to engage
young, non-technical, and underrepresented players in the world of
cyber security. \emph{P\&G} features a completely new design which
relies on probabilistic simulation to ensure fair gameplay.\emph{
}The game offers three major contributions. First, by exposing participant
to various types of cyber-attacks such as \emph{denial of service}
and \emph{masquerading attacks}, it builds cyber literacy in an inviting
way. Second, it teaches aspects of game theory such as information
asymmetry and deception implicitly. Finally, \emph{P\&G} engages players
with little or no previous introduction to cybersecurity. Indeed,
we conducted an initial set of tests with such a population at a community
center in the Lower East Side of Manhattan, New York. Encouraged by
these initial results, we hope to continue to improve \emph{P\&G}
so that it can contribute to the important and vast contemporary challenge
of cybersecurity education.

\end{document}